\documentclass{PoS}
\usepackage{cite}
\usepackage[utf8]{inputenc}

\newcommand{\lsim}{\raisebox{-0.07cm}{$\:\:\stackrel{<}{{\scriptstyle
 \sim}}\:\: $} }

\title{Implications of recent progress in heavy-quark hadro-production}

\ShortTitle{Implications of heavy-quark hadro-production}

\author{\speaker{Maria Vittoria Garzelli}
\thanks{Work supported by the Deutsche Forschungsgemeinschaft in Sonderforschungsbereich 676}
\\
        II. Institute for Theoretical Physics, University of Hamburg\\
        E-mail: \email{maria.vittoria.garzelli@desy.de}}

\author{Sven-Olaf Moch\\
        II. Institute for Theoretical Physics, University of Hamburg\\
        E-mail: \email{sven-olaf.moch@desy.de}}

\author{Guenter Sigl\\
        II. Institute for Theoretical Physics, University of Hamburg\\
        E-mail: \email{guenter.sigl@desy.de}}
\abstract{We discuss recent theoretical progress in heavy-quark hadro-production,
  in particular focusing on processes involving charm-quarks, and on their implications in different fields of particle phenomenology, from collider
  to astroparticle physics.}

\FullConference{12th International Symposium on Radiative Corrections (Radcor 2015) and LoopFest XIV (Radiative Corrections for the LHC and Future Colliders)\\
		15-19 June, 2015\\
		UCLA Department of Physics \& Astronomy Los Angeles, USA}

\begin{document}

\section{Introduction}
The production of heavy quarks, i.e., of quarks with a mass well above the QCD
scale parameter $\Lambda_{QCD}$,  was investigated since early days of QCD and
collider physics. The existence of a fourth quark, the charm, was theorized in
the sixties (see, e.g., Refs.~\cite{Bjorken:1964, Glashow:1970gm}) and confirmed
in 1974 by the observation of the J/$\Psi$ meson both at the Stanford Linear
Accelerator Center~\cite{Augustin:1974xw} and at the Brookhaven National
Laboratory~\cite{Aubert:1974js}. The existence of a third family of quarks was
postulated in 1973~\cite{Kobayashi:1973fv} and confirmed by the E288 experiment
at Fermilab, which observed bottomonium states~\cite{Herb:1977ek} in 1977, and
by the CDF and D0 collaborations at Tevatron, which in 1995 discovered the top
quark~\cite{Abe:1995hr, Abachi:1994td}. The main difference between the
relatively heavy (charm and bottom)  and really heavy top-quark relies on
their masses and on their lifetimes~\cite{Agashe:2014kda}. In particular, the
top-quark decays well before hadronizing, whereas charm and bottom, although
being, like the top, typically produced in perturbative processes (due to the
fact that, like $m_t$, even $m_c$ and $m_b \gg \Lambda_{QCD}$), are always
observed as intermediate hadronic states ($D$ and $B$ hadrons), and subsequently
decay in lighter ones eventually accompanied by prompt leptons. As a
consequence, the study of top-quark hadro-production, involving intermediate
top state reconstruction from $b$-jets and additional leptons, missing energy,
or light-jets, allows to get important information on the ``hard'' core of
hadron-hadron collision processes at energies above the top production
threshold, with hard-scatterings described by perturbation theory, and became 
possible only recently, i.e., at colliders with high enough center-of-mass
energies. On the other hand, studies of bottom and charm-quarks performed
since early days of accelerator physics with different beams 
have allowed not only for important tests of the validity of QCD and of the existence of new physics, but even to gain important information on the hadronization process~\cite{Norrbin:2000zc}.

Nowadays, most of the theoretical efforts at colliders concentrate on perturbative QCD, 
and the main focus of the heavy-quark working groups active at the most modern
collider experiment, the Large Hadron Collider (LHC) at CERN, is the study of
top-quark properties. In this respect, the hadronic pair-production of top-quarks 
is the most prominent signal, with tools for inclusive cross-section predictions 
up to NNLO QCD available~\cite{Aliev:2010zk, Czakon:2011xx}. 
At the differential level, first studies at NNLO for the
Tevatron~\cite{Czakon:2016ckf} and at the LHC~\cite{Czakon:2015owf} have appeared, 
and are actually necessary to meet the statistical accuracy reached during latest runs 
at the LHC. 
Additionally, recent theoretical developments have allowed to get
predictions beyond (N)LO also for different channels, ranging from single top
to $t\bar{t}$ hadro-production in association with other particles (one or more
vector bosons, the Higgs scalar, or jets). As for single top, NNLO QCD
corrections at the fully differential level have been computed for the
$t$-channel~\cite{Brucherseifer:2014ama}. On the other hand, as for processes
involving three or more final state particles at the parton level, the
state-of-the-art is represented by NLO QCD predictions matched to Parton
Shower (PS) approaches~\cite{Kardos:2011qa, Garzelli:2011vp, Garzelli:2012bn,
  Kardos:2014pba, Garzelli:2014aba, Cascioli:2013era, Hoeche:2014qda, Alwall:2014hca, Czakon:2015cla}, eventually accompanied by procedures for merging states
characterized by different light-jet multiplicities at the parton level~\cite{Frederix:2012ps, Gehrmann:2012yg}.
 Recently, additional efforts have also been devoted to the
inclusion of the electroweak (EW) corrections, first of all at fixed
order~\cite{Frixione:2014qaa, Frixione:2015zaa}, with general frameworks
matching NLO QCD + EW  corrections to PS and EW showers under development.

Although LHC is often considered as a top-quark factory~\cite{Han:2008xb}, one has
to take into account that the cross-sections for bottom and charm
hadro-production are even larger, making worth even the study of processes
involving the direct production of these quarks, without passing through
intermediate top states. In particular, for $pp$ collisions at $\sqrt{s}=13$~TeV,
the inclusive cross sections scale as 
$\sigma_{c\bar{c}} \sim \mathcal{O}(20) \sigma_{b\bar{b}} \sim  \mathcal{O}(20.000) \sigma_{t\bar{t}}$ 
at NNLO in QCD. Although selected
results from ATLAS and CMS are also available, charm and bottom-quark
hadro-production is being explored in particular at LHCb, which has provided
distributions for heavy meson and baryon hadro-production at the fully
differential level~\cite{Aaij:2013mga, Aaij:2015bpa}, together with results on
correlations in the production of heavy-meson pairs~\cite{Aaij:2012dz}. Charm
and bottom production measurements are useful for better understanding the
composition of the proton (an example is given in section 3), and the running
behavior of the strong coupling constant $\alpha_S$.

Furthermore, the experimental extraction of the values of the masses of heavy
quarks and their connection with theory, taking into account that quark masses
act as fundamental parameters in the Lagrangian of the Standard Model, has
attracted a lot of interest and efforts, but presents difficulties. In case of
top-quarks, long-standing discussions concern the relation between the
experimentally extracted mass and a rigorously defined theoretical
mass~\cite{Moch:2014tta}. On the other hand, in case of charm and bottom
quark, besides the complications inherent the description of the transition
between the perturbative regime and the non-perturbative one acting in the
hadronization, the main theoretical issue is represented by the slow
convergence of the perturbative series for quark masses in the on-shell
scheme. 
This is particularly serious in case of charm, for which recent calculations pointed
out the absence of convergence at 4-loop~\cite{Marquard:2015qpa}.

A recent review on top-quarks can be found, e.g., in
Ref.~\cite{delDuca:2015gca}. In the following, we will concentrate in
particular on the study of charm-quark hadro-production, by presenting some of
the implications of theoretical predictions and experimental results at LHC,
for astroparticle physics.

\section{Charm hadro-production cross-section}
\begin{figure}
\begin{center}
\includegraphics[width=0.48\textwidth]{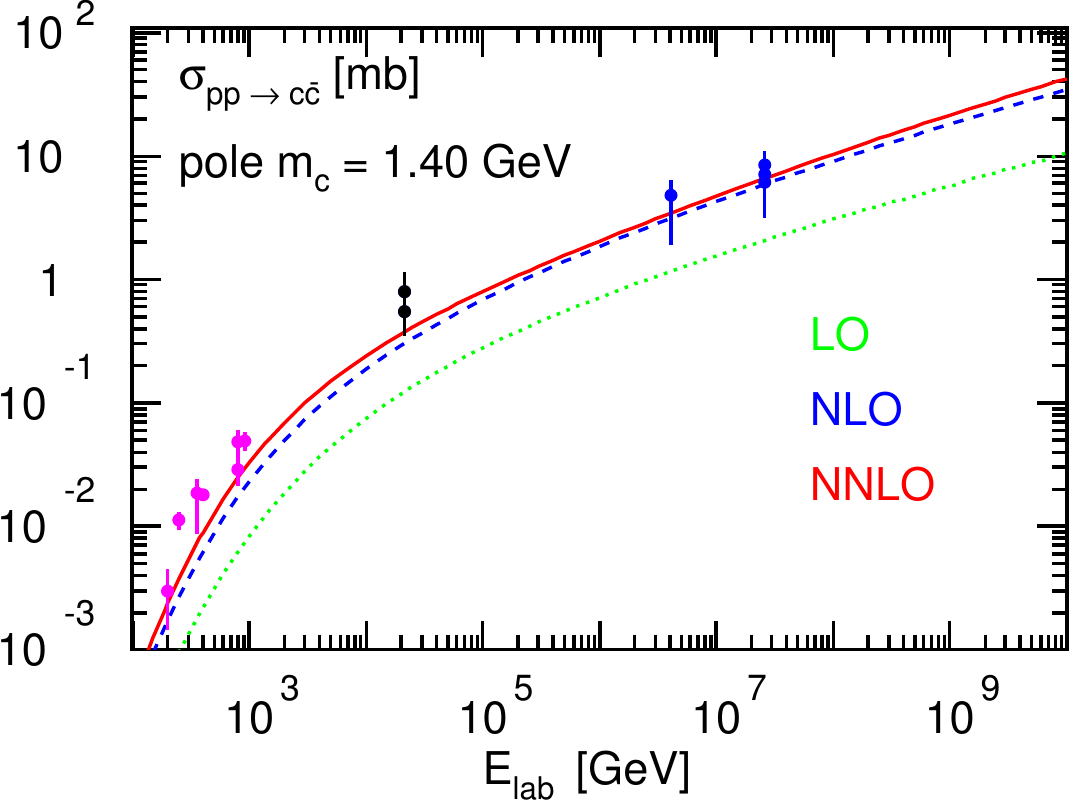}
\includegraphics[bb = 0 0 690 628, width=0.39\textwidth]{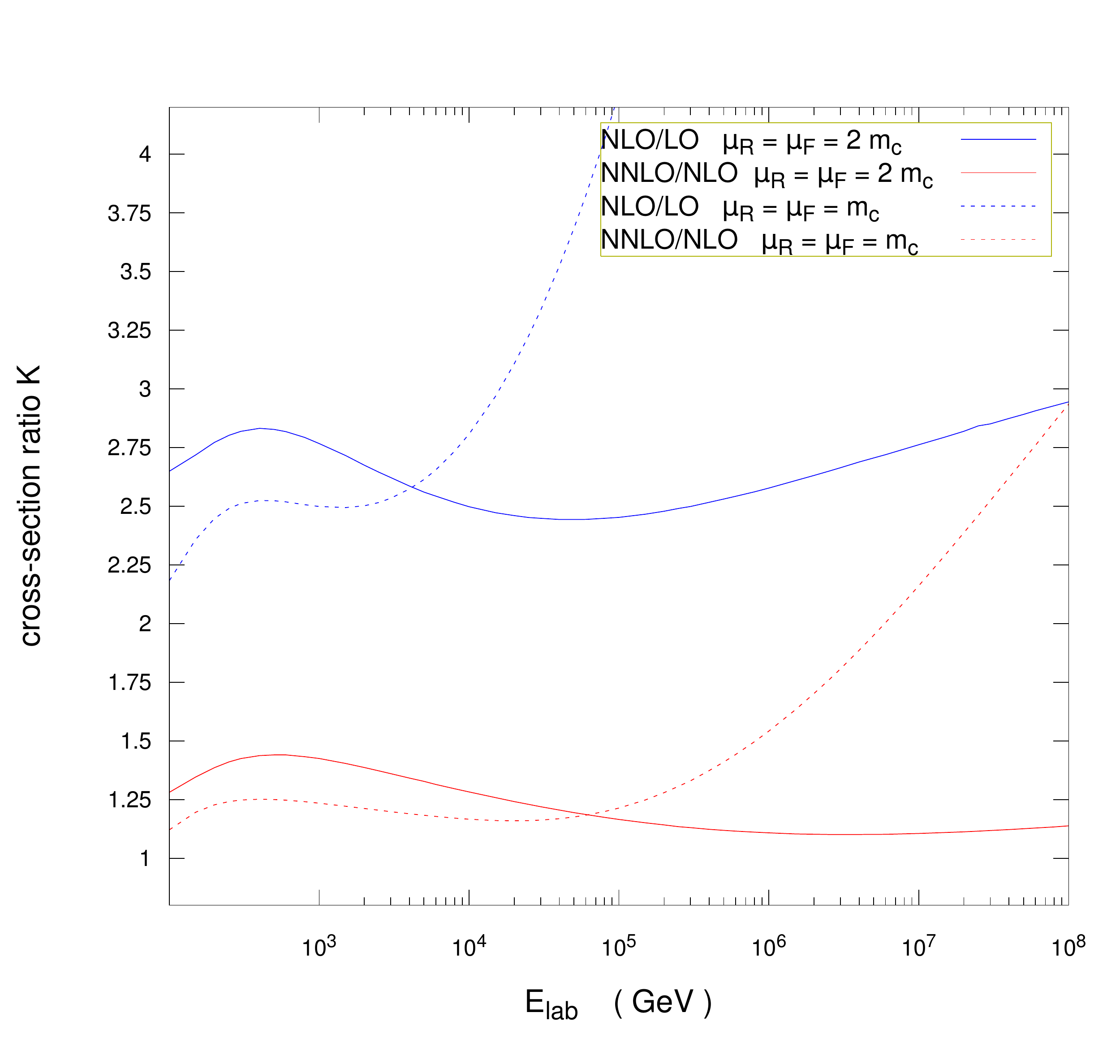}
\vspace{-0.1cm}\\
\caption{\label{figurak} 
  Total $\sigma_{pp \rightarrow c\bar{c}}$ as a
  function of the proton energy $E_p$ in the laboratory frame, including LO,
  NLO and NNLO QCD corrections, as compared to available experimental data
  from fixed target experiments~\cite{Lourenco:2006vw},
  RHIC~\cite{Adare:2006hc, Adamczyk:2012af}, ALICE~\cite{Abelev:2012vra},
  ATLAS~\cite{ATLAS:2011fea} and LHCb~\cite{Aaij:2013mga}. ABM11 PDFs at NNLO,
  $m_c$ = 1.4 GeV and $\mu_R = \mu_F = 2 m_c$  are used as input of the
  theoretical calculation in the fixed flavour number scheme with $N_f =
  3$. The corresponding NNLO/NLO and NLO/LO K-factors are shown in the right
  panel (solid lines), together with those corresponding to the scale choice
  $\mu_R = \mu_F = m_c$ (dotted lines).} 
\end{center}
\vspace{-0.65cm}
\end{figure}

Theoretical predictions for the total $c\bar{c}$ hadro-production cross-section
including NNLO QCD radiative corrections have been presented in
Ref.~\cite{Garzelli:2015psa} for a wide range of energies and compared to
experimental data from fixed target experiments, RHIC and LHC. These
predictions have been obtained by an extension of the {\texttt{Hathor}}
framework~\cite{Aliev:2010zk}, originally designed to compute $t\bar{t}$
hadro-production cross-sections. The comparison of the NNLO predictions with
the NLO ones turned out to show a good perturbative convergence even at
energies far larger that those reached at LHC, and allowed to estimate
$K$-factors related to NNLO/NLO ratios and to compare them with the more
commonly used $K$-factors for the NLO versus LO predictions. 
An example of typical cross-section values and related $K$-factors is shown in
Fig.~\ref{figurak}. The NNLO $K$-factor is smaller than the NLO one over the
whole range of considered energies for both the scale choices 
$\mu_R = \mu_F = m_c$ and $\mu_R = \mu_F = 2\, m_c$. Furthermore, while the first
choice produces smaller $K$-factors at lower energies, the second choice leads
to flatter $K$-factors, with a NNLO $K$-factor staying within 1.5 even at the
highest energy considered here ($E_p = 10^{10}$~GeV in the laboratory frame,
equivalent to about $E_{CM} \sim 137$~TeV in the center-of-mass reference
frame), where the $qg$(${\bar q}g$)-channels dominates over the other
initial-state partonic channels. 
The slope of the increase of the $K$-factor with $E_p$ at high
energy is shaped by the small-$x$ behavior of the parton luminosity and the
hard scattering cross sections. 
On the other hand, at low $E_p$, where all parton channels contribute at
medium to large $x$ significantly, the charm-quark mass $m_c$
is no longer negligible and the shape of both the NNLO and NLO $K$-factors is 
sensitive to charm production threshold effects~\cite{Kidonakis:2002vj}. 

\begin{figure}
\begin{center}
\includegraphics[width=0.45\textwidth]{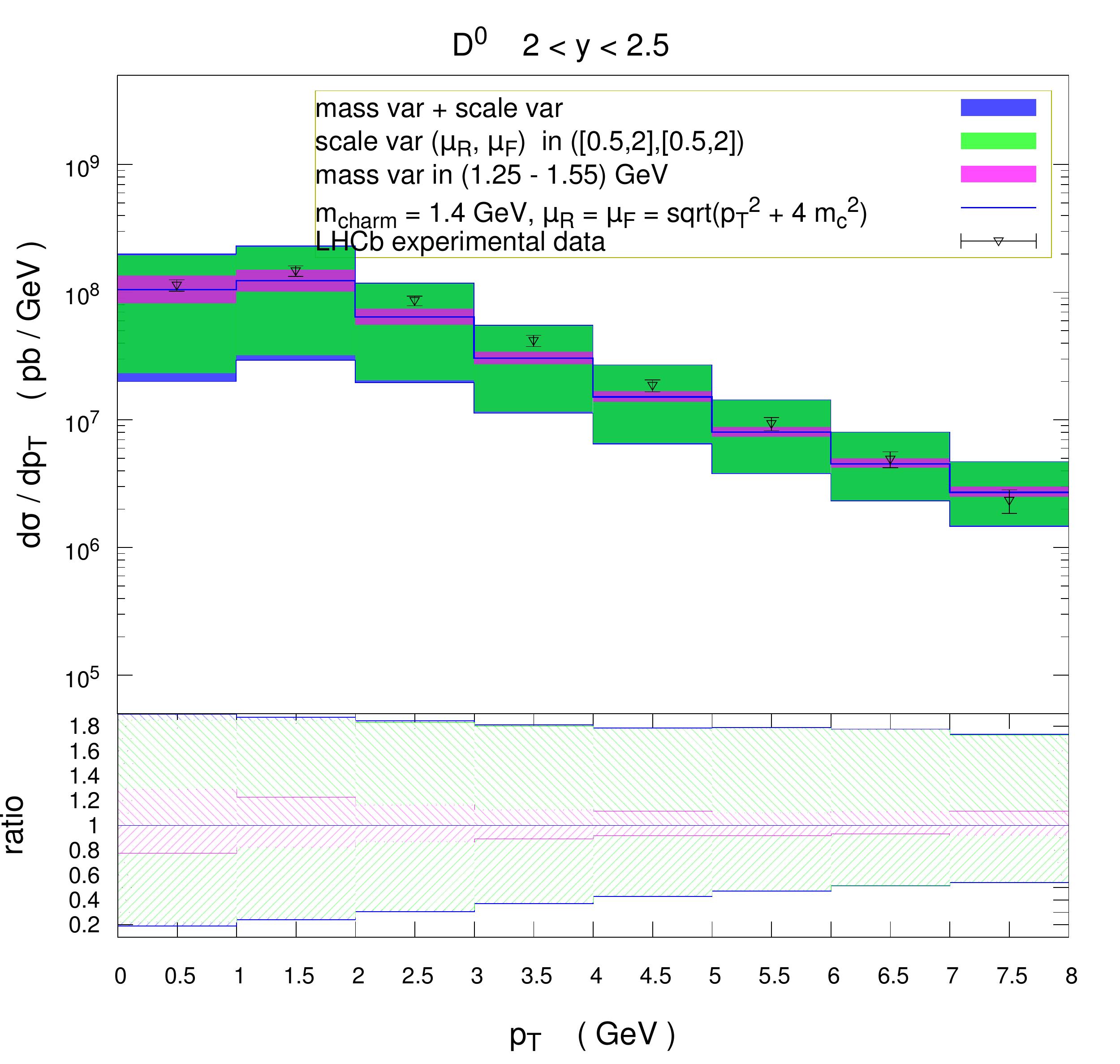}
\includegraphics[width=0.45\textwidth]{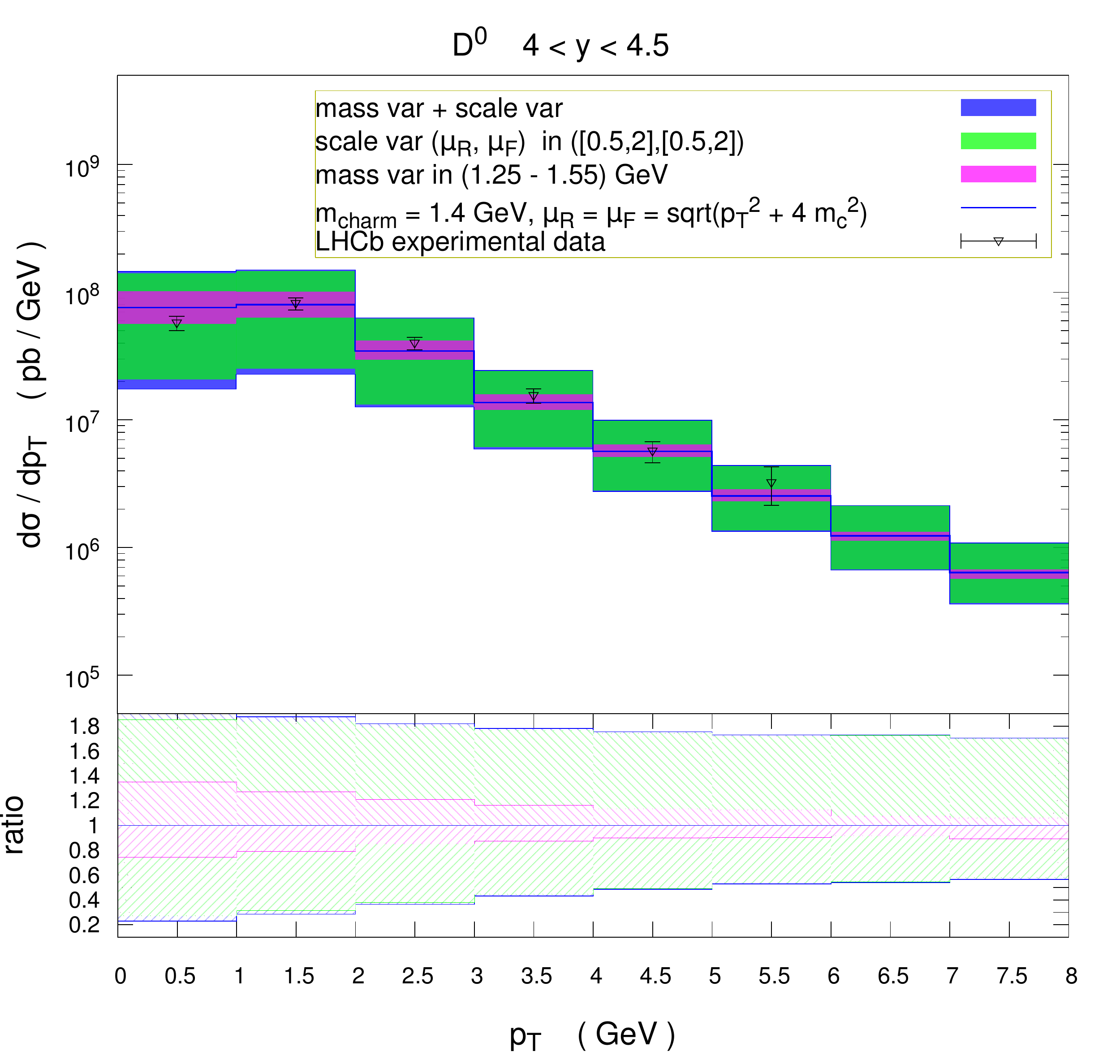}
\vspace{-0.1cm}\\
\caption{\label{figuralhcb} Theoretical predictions from \texttt{POWHEGBOX +
    PYTHIA6} for the $p_T$ distributions of $D^0$ mesons in $pp$ collisions at
  7 TeV vs. experimental data from LHCb. Scale and charm mass input are fixed
  as in Fig. 1.a. The NLO version of the ABM11 PDFs with
  $N_f=3$ is used. 
The violet and green band correspond respectively to charm mass and
  scale uncertainty, combined in quadrature in the blue band. The two panels
  correspond to the most central and the most peripheral rapidity bin explored
  at LHCb, respectively.}
\end{center}
\vspace{-0.65cm}
\end{figure}

At the differential level, predictions for theoretical distributions for
$c\bar{c}$ hadro-production are not yet available at NNLO. The state of the art
is represented in this case by NLO QCD approaches matched to PS. In some of
the available approaches and tools, EW corrections can be included as
well. However, taking into account that these corrections are in general
expected to be smaller than typical uncertainties due to renormalization and
factorization scale variation, we neglect electroweak effects in the following.
Theoretical predictions can be validated by comparison with experimental data from ALICE, ATLAS and LHCb. In particular, LHCb measured 
at both $\sqrt{s} = 7$~TeV and 13~TeV differential cross-sections for 
$D$-mesons in both transverse momentum and rapidity, considering the rapidity
range $2 \le y_0 \le 4.5$, 
corresponding to mid-peripheral collisions. 
An example of comparisons of the experimental data at 7 TeV with the
predictions by {\texttt{POWHEGBOX}}~\cite{Alioli:2010xd}~+~{\texttt{PYTHIA
    6.4.128}}~\cite{Sjostrand:2006za} using the ABM11~\cite{Alekhin:2012ig}
central PDF set at NLO, $\mu_R = \mu_F = \mu_0 = \sqrt{p_{T,c}^2 + 4 m_c^2}$ and a
charm-quark pole mass parameter $m_c = 1.4$~GeV (for further discussion concerning
the choice of these input see Ref.~\cite{Garzelli:2015psa}), is presented in
Fig.~\ref{figuralhcb}, showing consistency of theory with data even in the
largest rapidity bins: experimental data turned out to always lie within the
theory uncertainty bands due to scale and mass variation, obtained by
considering the interval $m_c \in [1.25 - 1.55]$~GeV and the 
($\mu_R$, $\mu_F$) combinations $\in$ [(2, 2), (0.5, 0.5), (2, 1), (1,~2), (0.5, 1),
(1,~0.5)] $\mu_0$. Similar agreement between data and theory was found even for
charged $D$-mesons.

\section{Implications of recent results of LHCb on Parton Distribution Functions}

Results of LHCb concerning charm and bottom hadro-production have been included
in recent PDF fits, due to the complementarity of their coverage in
Bjorken-$x$ with respect to that ensured by the data collected at HERA. In
particular the HERA charm data allowed to probe the gluon $x$ region
$10^{-4} \le x \le 10^{-1}$. On the other hand LHCb data, in particular those at
large rapidity ($\sim$~4~-~4.5)~allowed to extend the range on both sides,
giving rise to a total $x$  coverage extending to the interval $10^{-6} \lsim x \lsim  1$. 
This extension is especially important in case of high-energy
and very-high-energy collisions: in fact the higher are the energies, the more
asymmetric $g g$ initial states can come into play, characterized by 
$x_1 \ll 1$ and $x_2 \sim x_F$, with $x_F$ being the Feynman-$x$.

The first attempt to include LHCb data in PDF fits was performed by the PROSA
collaboration~\cite{Zenaiev:2015rfa}, using LHCb data at $\sqrt{s}$ = 7 TeV,
in association with data from HERA to perform the fit, following a procedure
similar to that originally used for performing the HERAPDF
fit~\cite{Aaron:2009aa}, further extended to the inclusion of hadron collider data. 
Both, the absolute LHCb data (i.e. $d^2\sigma$/$dp_T dy$) and the ratio of data
(d$\sigma$/$dy$)/($d\sigma/dy_0$) in each $p_T$ bin were fitted, with
$d\sigma/dy_0$ being the cross-section in the central bin $3 \le y_0 \le 3.5$
of the total measured rapidity range $2 \le y_0 \le 4.5$. 

The resulting PDF best-fits, together with the accompanying eigenvectors
corresponding to model, fit and parameterization uncertainties, extended PDF
grids down to $x \sim 10^{-6}$ and led to a reduced uncertainty for gluons in
the region $10^{-6} < x < 10^{-4}$ with respect to previous fits. 

A second attempt was performed by members of the NNPDF collaboration, using
LHCb data at $\sqrt{s}=7$~TeV and 13 TeV to further constrain the most recent NNPDF fit. In
particular, it turned out that the NNPDF3.0 fit~\cite{Ball:2014uwa} gives rise
to uncertainty bands which open up dramatically for $x < 10^{-4}$, where
NNPDF3.0 PDFs are essentially unconstrained. Including LHCb data has led to a
new fit, labeled as NNPDF3.0~+~LHCb~\cite{Gauld:2015yia}, with modifications
of the shape of the central gluon density as a function of $x$ and reduced
uncertainty band in the low $x$ region. Although this fit is supposed to
represent an improvement with respect to the NNPDF3.0 PDF version nowadays
widely used at colliders, it is not yet publicly available, at least as far as
we know.  

We also notice that the ABM11 and ABM12~\cite{Alekhin:2013nda} fits, although
having been done well before and thus non including any LHCb data, for low-$x$
gluons turned out to have all their eigenvectors well included within the
uncertainty bands of the PROSA fit and to present compatibility even with the
NNPDF3.0~+~LHCb uncertainty band.

\section{Astrophysical implications of charm hadro-production}
\begin{figure}
\begin{center}
\includegraphics[width=0.245\textwidth]{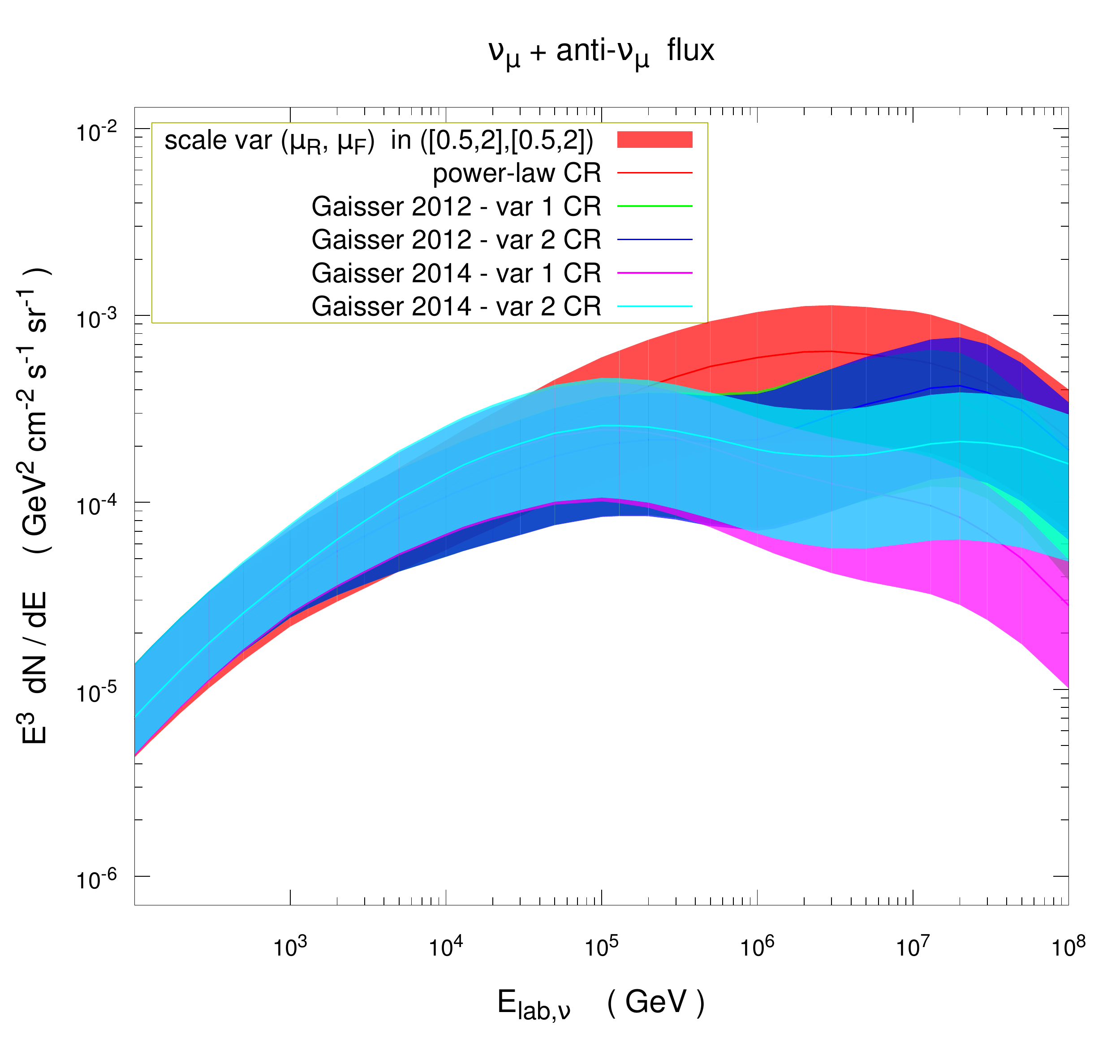}
\includegraphics[width=0.245\textwidth]{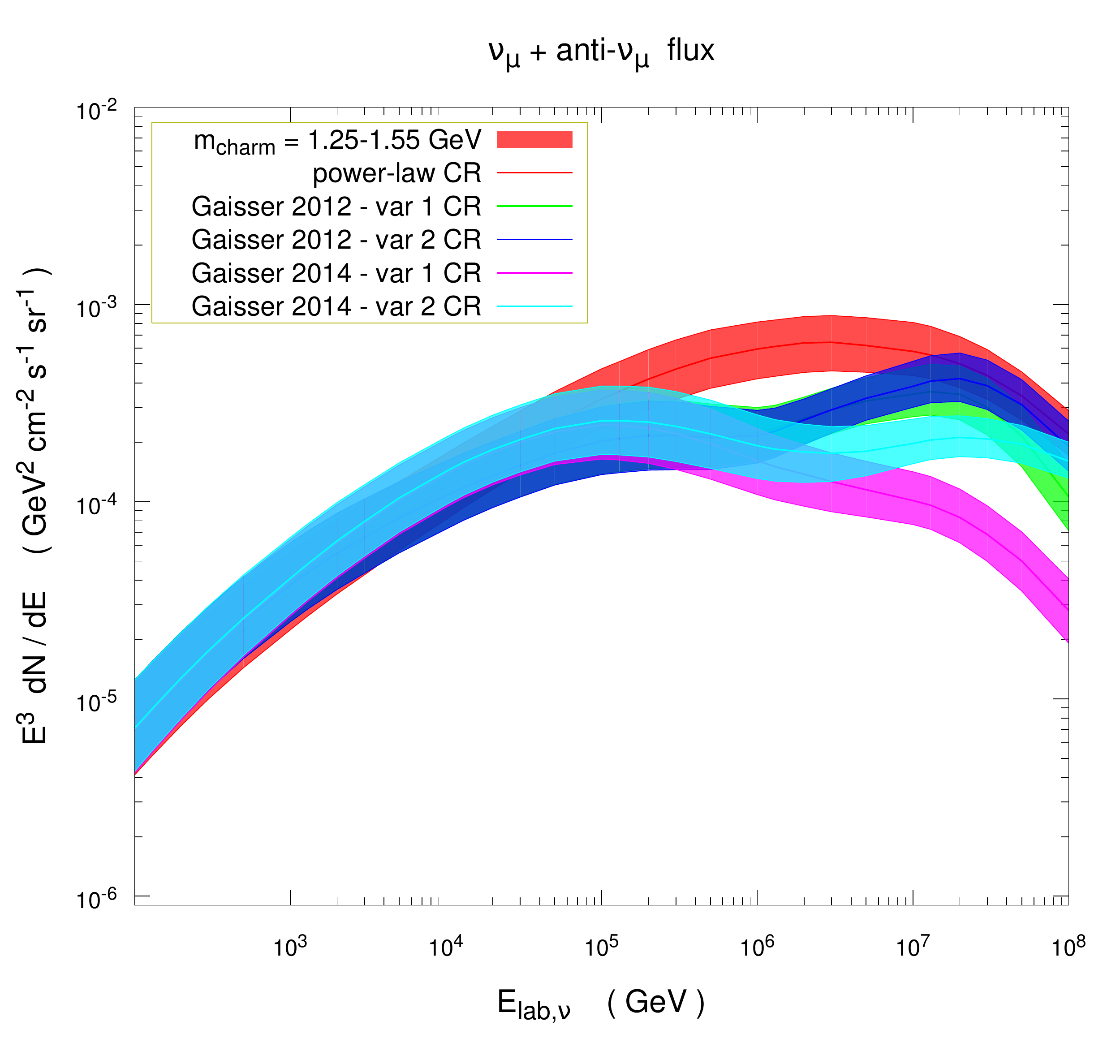}
\includegraphics[width=0.245\textwidth]{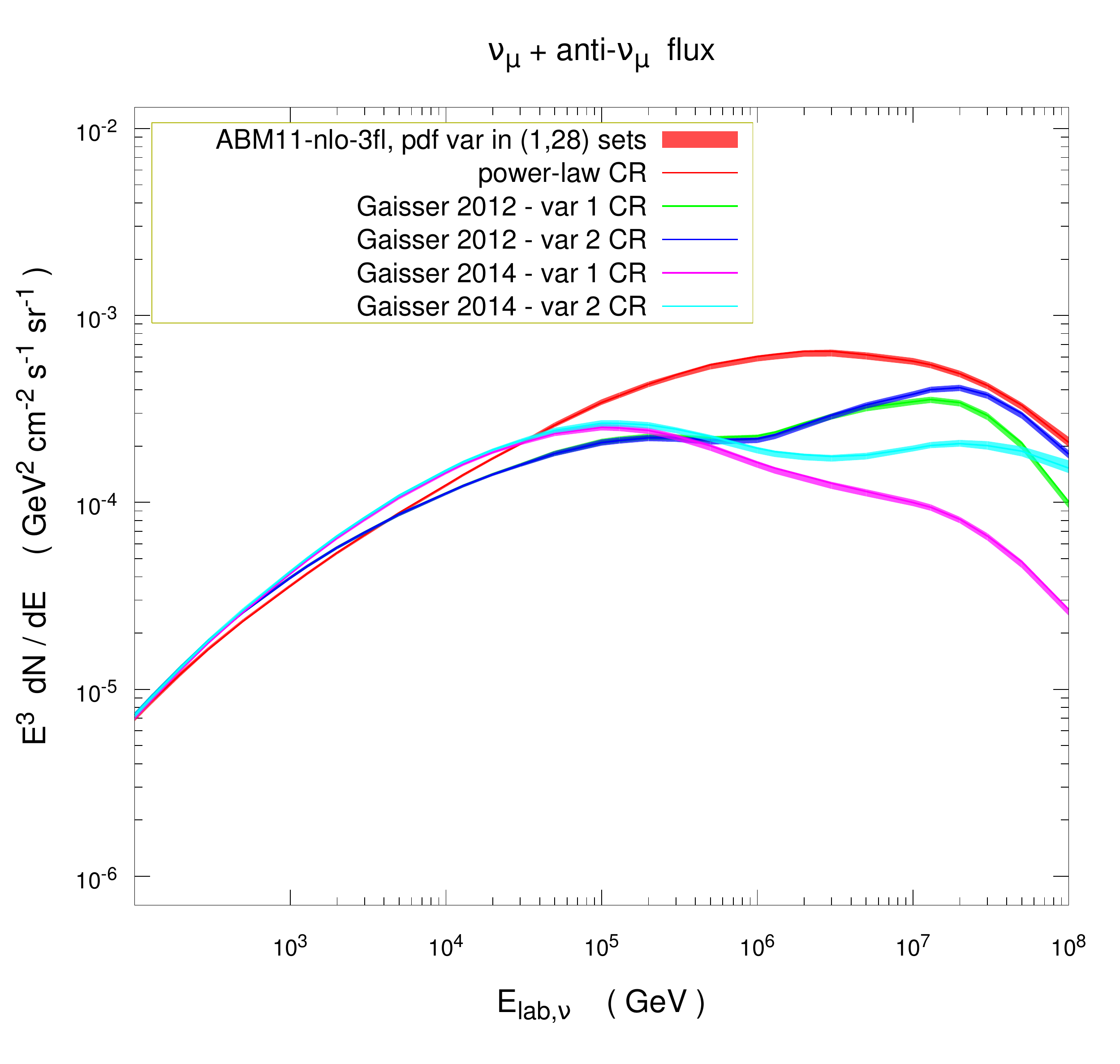}
\includegraphics[width=0.245\textwidth]{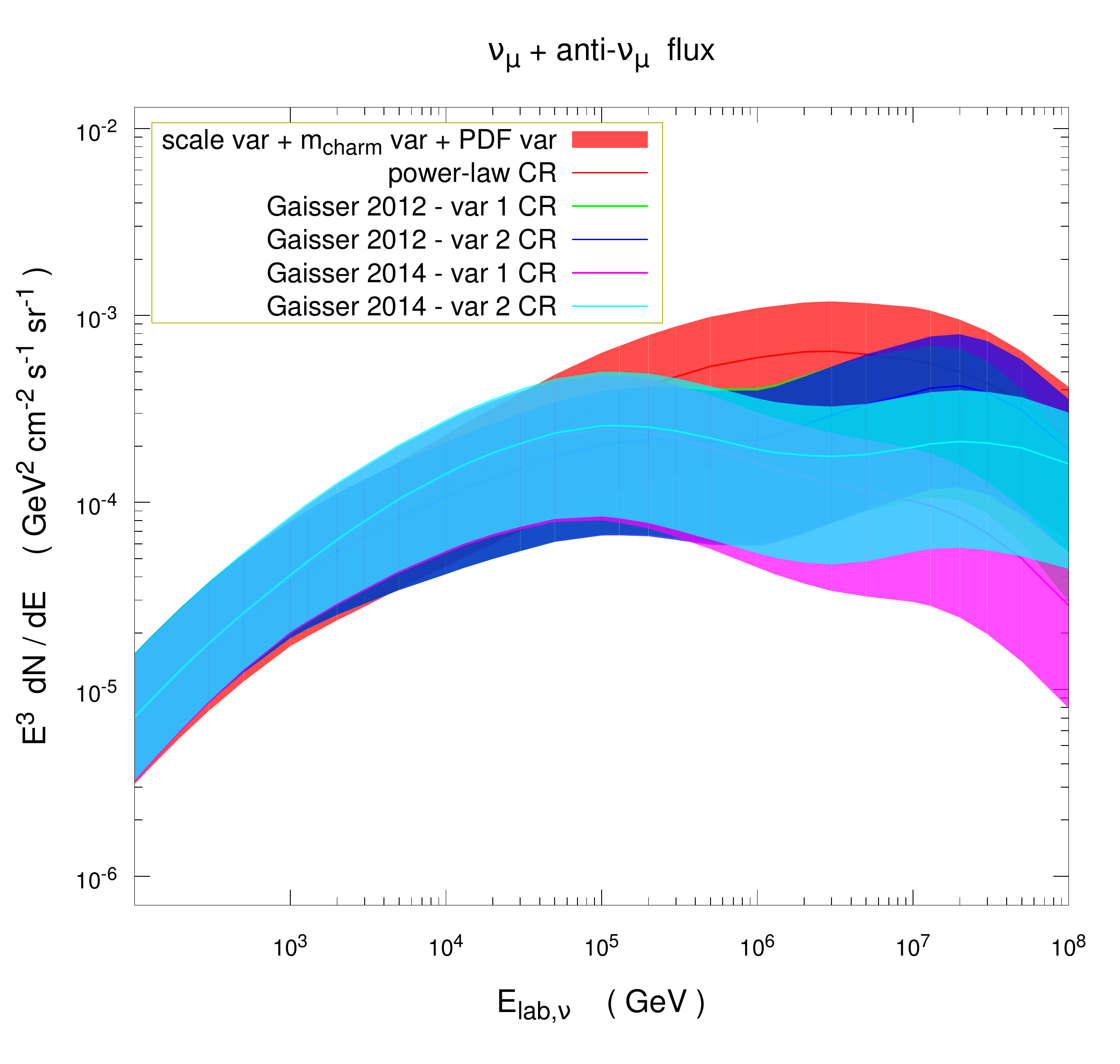}
\vspace{-0.1cm}\\
\caption{\label{figflux} Prompt ($\nu_\mu$ + $\bar{\nu}_\mu$) fluxes as a function of $E_\nu$ for five different primary CR input fluxes: uncertainties due to $\mu_R$ and $\mu_F$ scale variation, $m_c$ variation and PDF variation (limited to the 28 eigenvectors of the ABM11 NLO PDF fit), together with their combination in quadrature are shown in panel a, b, c, d, respectively. QCD inputs are chosen as in Fig.~2. Predictions available at {\texttt{http://www.desy.de/$\sim$lepflux}}.}
\end{center}
\vspace{-0.65cm}
\end{figure}

The considerations made above on charm hadro-production in QCD find an
interesting application and test-bed in astroparticle physics, in the problem
of the evaluation of prompt neutrino fluxes~\cite{Garzelli:2015psa}. In
particular, it is well known that the bulk of atmospheric neutrinos is due to
the leptonic and semi-leptonic decay of light mesons (mostly $\pi^\pm$'s and
$K^\pm$'s) originated by the interaction of primary cosmic rays (CR) impinging
into the Earth's atmosphere with the atmospheric
nuclei~\cite{Lipari:1993hd}. This conventional neutrino flux,
peaked at neutrino energies $E_\nu$ below 1~GeV and known to fall down
quite rapidly with  $E_\nu$, is investigated since many years~(see
e.g., Ref.~\cite{Honda:2006qj}). On the other hand, at higher energies, the
IceCube collaboration has recently reported evidence for a leptonic flux
extending at least up to the $\sim$ PeV energy region~\cite{Aartsen:2013jdh,
  Aartsen:2014gkd}, whose origin is still under discussion. Although it is
believed that a large portion of this flux has an astrophysical origin,
i.e., it is due to neutrinos reaching the Earth from deep space, traveling
almost undeflected from far galactic or extra-galactic sources, a rigorous
investigation has to consider and quantify which fraction of these neutrinos
can instead originate in the Earth's atmosphere. The latter process would be
possible at energies like those explored by IceCube, thanks to semi-leptonic
decays of $D$-mesons and baryons, eventually formed in the interaction of CR
primaries with the atmosphere. 
At energies high enough this process is
more effective in producing neutrinos than the process of conventional
production described above because highly
boosted $\pi^\pm$ and $K^\pm$ have a suppressed probability to decay in
neutrinos, while crossing the Earth's atmosphere,
whereas semi-leptonic decays
would still be abundant for $D$-mesons and baryons, having a larger mass, a
smaller life-time ($\tau_{\, 0,\,D} \sim 10^{-22}$ s versus 
$\tau_{\, 0,\,\pi} \sim 10^{-8}$ s) and decaying just immediately after their
formation. This neutrino component is thus called prompt component. We have 
computed it in QCD, by describing CR interactions with the Earth's atmosphere in
terms of nucleon-nucleon ($NN$) collisions and by using the same tools and input
described in Section 2 as for the simulation of the $NN$ $\rightarrow$
$c\bar{c}$ $\rightarrow$ $D + X$  processes. In particular, predictions from
{\texttt{POWHEGBOX}}~+~{\texttt{PYTHIA6}} in the same configuration validated
with respect to LHCb data at 7 TeV (see Fig. 1) were used even to
compute the prompt neutrino flux presented in our
paper~\cite{Garzelli:2015psa}. More recently another group adopted a similar
framework ({\texttt{POWHEGBOX}} + {\texttt{PYTHIA8}}~\cite{Sjostrand:2007gs})
together with the NNPDF3.0~+~LHCb PDFs, getting predictions for prompt
neutrino fluxes as well~\cite{Gauld:2015kvh}, with a method very close to our
one~and~same~CR~primary~input~fluxes.  

\begin{figure}
\begin{center}
\includegraphics[width=0.45\textwidth]{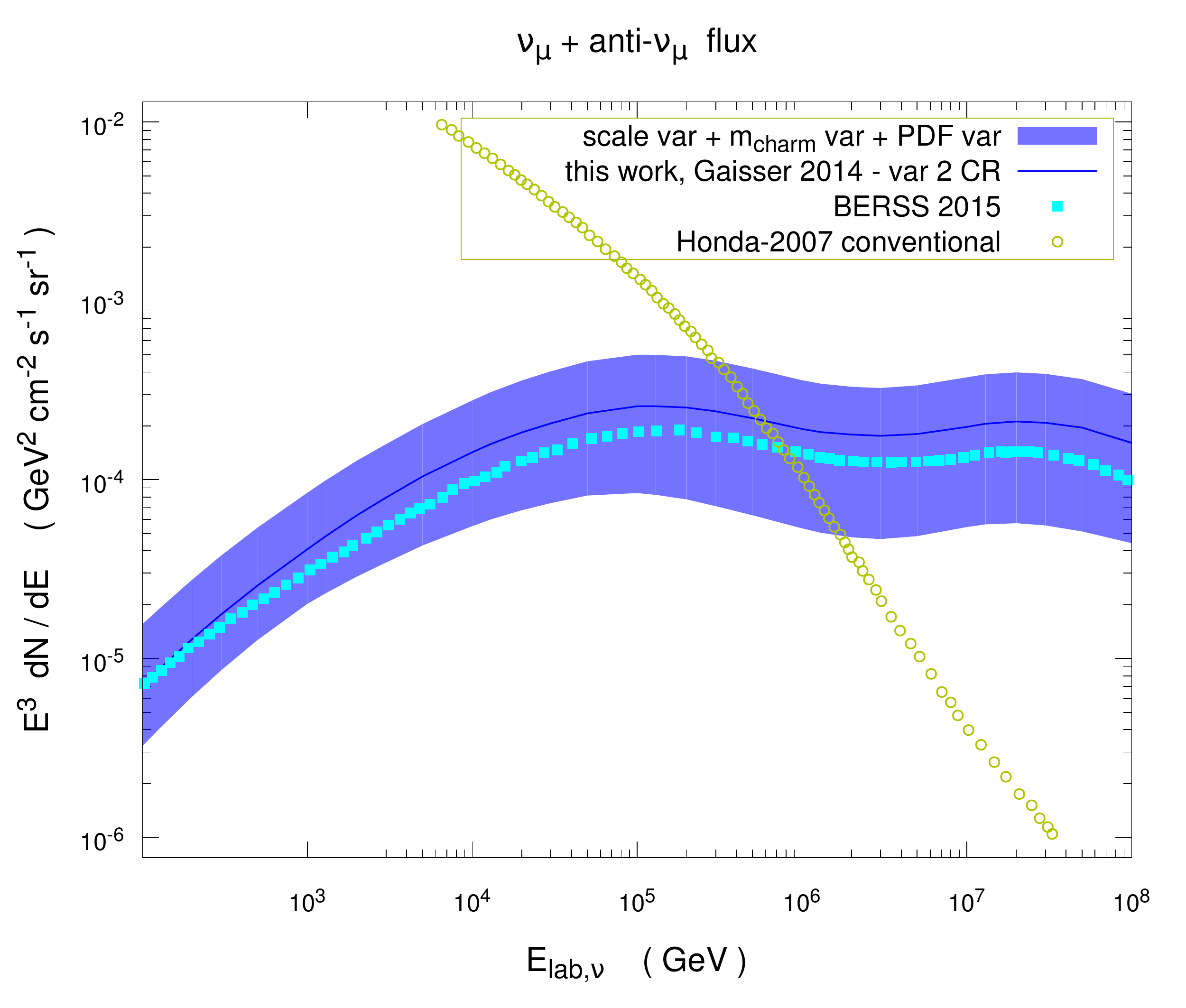}
\includegraphics[width=0.403\textwidth]{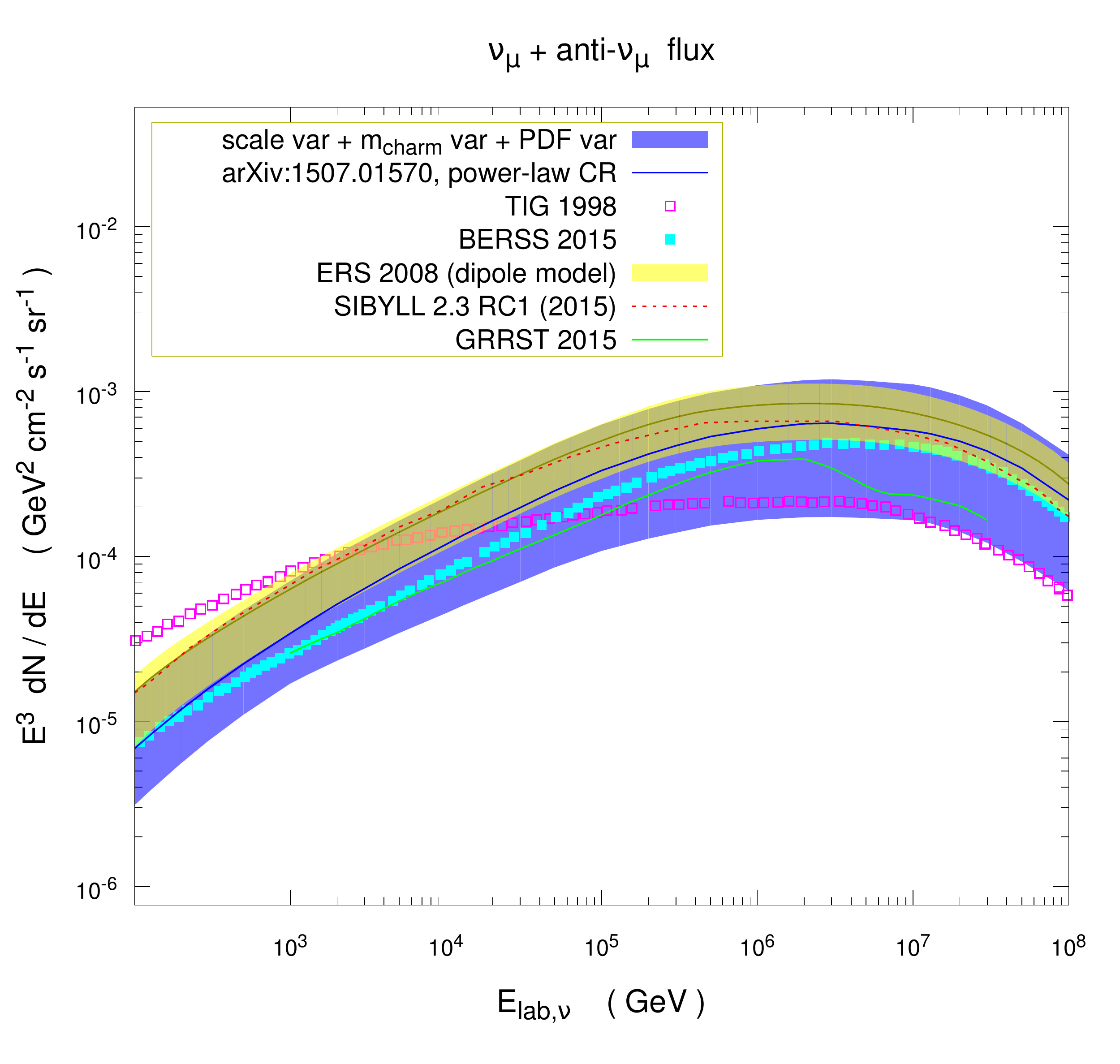}
\vspace{-0.1cm}\\
\caption{\label{figtrans} a) prompt ($\nu_\mu$ + $\bar{\nu}_\mu$) flux and its
  uncertainty vs. conventional flux for one of the Gaisser primary CR spectrum
  variants; b) a comparison between our predictions for the prompt flux (blue line and bands) and other recent predictions for the power-law CR primary spectrum. See text for more detail. }
\end{center}
\vspace{-0.65cm}
\end{figure}

Our predictions, together with separate uncertainty bands due to scale, charm
mass and PDF variation (the latter restricted to the 28 eigenvectors of the
ABM11 NLO PDF fit), are shown in Fig.~\ref{figflux}, for five different CR primary fits~\cite{Gaisser:2013bla, Stanev:2014mla}. Actually, it is evident that at
high energy, the astrophysical uncertainty related to our poor knowledge of
primary CR fluxes becomes even more important than the QCD uncertainty. This
means that if, on the one hand, these predictions can be further refined by a
future improved QCD description of charm hadro-production including effects
from higher order corrections capable of limiting/overcoming the huge scale
uncertainty of the NLO computation, amounting to many ten percent, on the
other hand, it is also indispensable to improve our knowledge of CR fluxes by
means of forthcoming CR data from extended air shower experimental arrays,
like e.g. the Pierre Auger Observatory.

The relative importance of conventional and prompt neutrino flux is presented
in Fig.~\ref{figtrans}.a, showing a transition energy, i.e. the energy where
the prompt flux overcomes the conventional one, around $E_{\nu, \, trans} = 6^{\, + 12}_{\, - 3} \cdot 10^5$~GeV. 

Finally, a comparison of our predictions with other old~\cite{Gondolo:1995fq, Enberg:2008te} and recent predictions~\cite{Bhattacharya:2015jpa, Fedynitch:2015zma, Gauld:2015kvh} obtained during last year, is presented in Fig.~\ref{figtrans}.b, showing that all most recent predictions considered turn out to lie within our theoretical uncertainty band.



\providecommand{\href}[2]{#2}\begingroup\raggedright\endgroup


\end{document}